\documentclass[11pt,a4paper]{article}

\usepackage{xcolor}
\usepackage[left= 2cm, right= 2cm]{geometry}
\usepackage{amsmath,latexsym,amssymb,slashed}
\usepackage{graphicx}				
\usepackage{amssymb}					
\usepackage{tabularx}				
\usepackage[vcentermath]{youngtab}	

\usepackage[latin1]{inputenc}
\usepackage[T1]{fontenc}
\usepackage{slashed}

\newcommand{\eq}[1]{\begin{equation}#1\end{equation}}
\newcommand{\spl}[1]{\begin{split}#1\end{split}}

\def\bea{\begin{eqnarray}}
\def\eea{\end{eqnarray}}

\newcommand{\boxedeq}[1]{
\begin{equation}
\fbox{
\rule[0.7cm]{0pt}{0pt}
$#1$
\rule[-0.45cm]{0pt}{0pt}
}
\end{equation}
}

\def\d{\text{d}}

\newcommand{\swed}{{\scriptscriptstyle \wedge}}

\begin{document}
\setlength{\parindent}{0pt}

\begin{titlepage}
	\vspace{14pt}
	
	\begin{center}
		
		{\Large \bf 
		Consistent truncation and de Sitter space\\ from  gravitational instantons }\\

		\vspace{1.6cm}
		
		{\bf \footnotesize Robin Terrisse\footnote{terrisse@ipnl.in2p3.fr, $^2$tsimpis@ipnl.in2p3.fr} and Dimitrios Tsimpis$^{2}$ }
		
		\vspace{1cm}
		
		{\bf }

		\vspace{.2cm}
		
		{\it   Institut de Physique Nucl\'eaire de Lyon  }\\
		{\it Universit\'e de Lyon, UCBL, UMR 5822, CNRS/IN2P3 }\\
		{\it 4 rue Enrico Fermi, 69622 Villeurbanne Cedex, France  }\\

		\vspace{14pt}

	\end{center}
	\begin{abstract}
	\noindent	
We construct a four-dimensional consistent truncation to the bosonic part of the universal sector of Calabi-Yau IIA compactification (i.e.~the gravity multiplet, one vectormultiplet, and one hypermultiplet) in the presence of background  flux and fermionic condensates generated by gravitational instantons. 
The condensates are controlled by the ratio of the characteristic length of the Calabi-Yau to the string length, and can be fine-tuned to be dominant in a region of large volume and 
small string coupling. The consistent truncation admits de Sitter solutions supported by the condensates, subject to certain validity conditions that we discuss. 

	\end{abstract}	
	
\end{titlepage}


\tableofcontents


\newpage

\section{Introduction}

\setcounter{footnote}{0}

Within the framework of critical ten-dimensional superstring theories, it has proven very difficult to obtain controlled scenarios leading to a low-energy effective theory with a positive cosmological constant (CC) or dark energy, see \cite{Danielsson:2018ztv,Palti:2019pca} for recent reviews. On rather general grounds, the presence of de Sitter vacua within the limit of classical two-derivative supergravity is excluded \cite{deWit:1986mwo,Maldacena:2000mw}, thus one is led to include quantum corrections in order to evade the no-go theorem.\footnote{The no-go theorem might also be circumvented at the classical level by including orientifold sources (see \cite{Cordova:2018dbb} for a recent proposal), although this scenario is also subject to stringent constraints \cite{Andriot:2018ept,Cribiori:2019clo,Andriot:2019wrs}.} One such quantum effect is fermionic condensation, which is known to occur in supersymmetric Yang-Mills theories \cite{Novikov:1983ee}.

The vacuum expectation value (VEV) of all single fermions should vanish in order to preserve the symmetries of a maximally-symmetric vacuum of the theory,  however non-vanishing quadratic or quartic fermion VEV's  (condensates) may still be generated by non-perturbative effects such as instantons. 
Fermionic condensates have thus the potential to generate a positive contribution to the CC. Within the framework of the ten-dimensional superstrings, most studies have focused on gaugino condensation in the heterotic theory \cite{Dine:1985rz, Derendinger:1985kk, LopesCardoso:2003sp, Derendinger:2005ed, Manousselis:2005xa, Chatzistavrakidis:2012qb, Gemmer:2013ica, Minasian:2017eur}, which however does not seem to allow a positive CC \cite{Quigley:2015jia}. Recent results \cite{Soueres:2017lmy,Terrisse:2018qjm} indicate that the situation is more encouraging within the framework of the IIA superstring --which appears to allow the generation of a positive CC by fermionic condensates, at least in principle.

In the functional integration over metrics approach to quantum gravity \cite{Gibbons:1978ac}, the gravitino condensates arise from saddle points of the 4d action corresponding 
to gravitational instantons. These are noncompact asymptotically locally Euclidean (ALE) spaces with self-dual Riemann curvature and thus vanishing Einstein action \cite{Eguchi:1980jx}. 
Although other saddle points may exist in the presence of matter fields \cite{Witten:1981nf}, they would have positive Euclidean  action.\footnote{This follows from the positive action conjecture \cite{Gibbons:1978ac}, which in its turn can be seen to follow from the positive energy theorem \cite{Schon:1979uj,Schon:1979rg,Schon:1981vd,Witten:1981mf}.} Thus ALE spaces are expected to capture the dominant instanton contributions in the path integral approach of quantum gravity.

Going beyond the two-derivative approximation of the 4d effective action, the gravitational instantons give a positive contribution to the action at the four-derivative order. 
Among the ALE spaces, the one with the minimal four-derivative action  is the Eguchi-Hanson (EH) gravitational instanton \cite{Eguchi:1978gw}. 
In the EH background there are two positive-chirality spin-3/2 zero modes of the Dirac operator, and no spin-1/2 zero modes, thus giving rise to a nonvanishing gravitino bilinear condensate in 
four dimensions at one loop in the 4d gravitational coupling \cite{Hawking:1979zs,Konishi:1988mb}. The quartic gravitino VEV's receive contributions from 
 ALE instantons with higher Hirzebruch signature. Since ALE spaces do not support spin-1/2  zero modes, no dilatino condensates are generated.

The aim of the present paper is to study  gravitino condensation in the IIA theory compactified on Calabi-Yau (CY) threefolds.  To that end we construct a 4d consistent truncation capturing the bosonic part of the universal sector of CY IIA compactification (i.e.~the gravity multiplet, one vectormultiplet, and one hypermultiplet) in the presence of background  flux and gravitino condensates generated by ALE instantons.

In the limit of vanishing flux and condensates,  our construction reduces to the universal bosonic sector of the effective action of IIA CY compactifications (at the two-derivative order) thus proving that the latter is also a consistent truncation. 
In the presence of nonvanishing flux and fermion condensates, the result should  be thought of as 
a subsector of the 4d effective action, in the limit where the masses induced by the flux and/or the condensate are sufficiently smaller than the Kaluza-Klein (KK) scale.

The condensates are controlled by the ratio of the characteristic length of the CY to the string length, and can be fine-tuned to be dominant in a region of large volume and 
small string coupling. The consistent truncation admits de Sitter solutions supported by the condensates, subject to certain validity conditions that we discuss.

The plan of the remainder of the  paper is as follows. Section \ref{sec:review} includes a brief review of the reduction of IIA on CY at the two-derivative level, in the absence of flux and condensates. Higher-order derivative corrections in the 4d action are discussed in section \ref{sec:dercorrections}. 
The consistent truncation to the universal sector is contained in section \ref{sec:ct}: 
 in section \ref{sec:consistent} we construct the truncation in the presence of background flux. A further extension to 
include gravitino condensates  is constructed in section \ref{sec:fermcond}.  Maximally-symmetric vacua thereof are discussed in section \ref{sec:vacua}. 
Section \ref{sec:discussion} discusses the conditions of validity of 
our results, and some open directions. 
Appendix \ref{app:spin}  discusses our spinor conventions.  
A brief review of ALE spaces is 
given in section \ref{app:ale}.  The general form of the gravitino condensates, generated by ALE gravitational instantons in the context of 4d $\mathcal{N}=1$ supergravity, is reviewed in appendix \ref{app:grcond}.

\section{Review of IIA reduction on CY}\label{sec:review}

To establish notation and conventions, let us  briefly review the reduction of IIA on CY at the two-derivative level, in the absence of flux and condensates. 
As is well known, the KK reduction of (massless) IIA supergravity around the fluxless $\mathbb{R}^{1,3}\times Y$ vacuum results in a 4d $\mathcal{N}=2$ supergravity, 
 whose bosonic sector  consists of one gravity multiplet (containing  the metric and one vector), $h^{1,1}$ vector multiplets (each of which consists of one vector and two real scalars) and $h^{2,1}+1$  hypermultiplets (each of which contains four real scalars), where $h^{p,q}$ are the Hodge numbers of the CY threefold $Y$. 
The $2h^{1,1}$ real scalars $(v^A,\chi^A)$ in the vector multiplets come from the NS-NS $B$ field and deformations of the metric of the form,
\eq{\label{4}
B=\beta(x)+\sum_{A=1}^{h^{1,1}}\chi^A(x)e^A(y)
~;~~~
i\delta g_{a\bar{b}}= \sum_{A=1}^{h^{1,1}}  v^A(x)e^A_{a\bar{b}}(y)
~,}
where $\beta$ is a two-form in $\mathbb{R}^{1,3}$;  $\{e^A_{a\bar{b}}(y),~A=1,\dots,h^{1,1}\}$ is a basis of harmonic (1,1)-forms on the CY, and $x$, $y$ are coordinates 
of $\mathbb{R}^{1,3}$, $Y$ respectively; 
 we have introduced holomorphic, antiholomorphic  internal indices from the beginning of the latin alphabet: $a=1,\dots,3$, $\bar{b},=1,\dots,3$, respectively. 
Since every CY has a K\"{a}hler form (which can be expressed as a linear combination of the basis (1,1)-forms), there is a always at least one vector multiplet (which may be called ``universal'', in that that it exists for any CY compactification) whose scalars consist of the volume modulus $v$ and one scalar $\chi$.

The $2(h^{2,1}+1)$ complex scalars of the hypermultiplets, and the $h^{1,1}+1$ vectors of the gravity and the vectormultiplets arise as follows: from the one- and three-form RR potentials $C_1$, $C_3$ and the complex-structure deformations of the metric,\footnote{The right-hand side of the first equation of \eqref{7} can be seen to be automatically symmetric in its two free indices.}
\eq{\spl{\label{7}
\delta g_{\bar{a}\bar{b}}&=\sum_{\alpha=1}^{h^{2,1}} \zeta^{\alpha}(x) \Omega^{*cd}{}_{\bar{a}}
\Phi^{\alpha}_{cd\bar{b}}(y)~;~~~C_1=\alpha(x)~;
\\
C_3&=-\frac12\Big(\xi(x)\text{Im}\Omega+\xi'(x)\text{Re}\Omega\Big)
+\sum_{A=1}^{h^{1,1}}\gamma^A(x)\wedge e^A(y)
+\Big(\sum_{\alpha=1}^{h^{2,1}} \xi^\alpha(x) \Phi^\alpha(y)+\mathrm{c.c.}\Big)
~,}}
where $\Omega(y)$ is the holomorphic threeform of the CY and  $\{\Phi^{\alpha}_{ab\bar{c}}(y),~\alpha=1,\dots,h^{2,1}\}$ is basis of harmonic (2,1) forms  on the CY, 
we obtain the complex scalars $(\zeta^\alpha, \xi^\alpha)$ and the vectors $(\alpha, \gamma^A)$. Moreover the real scalars $(\xi,\xi')$ together with the dilaton $\phi$ and the axion $b$ combine into one universal hypermultiplet. Recall that  if $h$ is the 4d component of the NSNS three-form, 
\eq{ h=\d \beta~,}
the axion $b$ is given schematically by $\d b \sim\star_4 h$ (the precise relation is eq.~\eqref{prhd} below).

In summary, 
the universal bosonic sector of the 4d $\mathcal{N}=2$ supergravity arising from IIA compactification on $Y$ contains 
the metric and the vector of the gravity multiplet $(g_{\mu\nu},\alpha)$, the vector and the 
the scalars of one  
vectormultiplet $(\gamma,v,\chi)$,  and the scalars of the universal 
hypermultiplet $(\xi,\xi',\phi,b)$.

\subsection{Derivative corrections}\label{sec:dercorrections}

Four-derivative corrections to the 4d effective action 
resulting from compactification of the IIA superstring on CY threefolds 
have been known since \cite{Antoniadis:1997eg}. More recently they have been computed in \cite{Grimm:2017okk} (see also \cite{Weissenbacher:2019mef}) from compactification of 
certain known terms of the ten-dimensional IIA tree-level and one-loop superstring effective action at order $\alpha^{\prime3}$.  
The authors of that reference take into account the graviton and $B$-field eight-derivative terms given  in \cite{Gross:1986mw, Liu:2013dna}, but neglect e.g. the dilaton 
derivative couplings and RR couplings of the 
form $R^2(\partial F)^2$ and $\partial^4F^4$ calculated in \cite{Policastro:2006vt}. Furthermore \cite{Grimm:2017okk} neglects loop corrections from massive KK fields.\footnote{Presumably the KK loop corrections are subleading and vanish in the large-volume limit (see however \cite{Haack:2015pbv} for an exception to this statement). At any rate these corrections are dependent on the specific CY and at the moment can only be computed on a case-by-case basis, e.g.~around the orbifold limit where the CY reduces to $T^6/\Gamma$ with $\Gamma$ a discrete group. Winding modes are heavier than KK modes in a regime where \eqref{eqlim} holds.} 

In a low-energy expansion, the 4d effective action takes the schematic form \cite{Katmadas:2013mma},
\eq{\label{scm}
2\kappa^2S=\int\d x^4\sqrt{g}\left( 
R
+\beta_1\alpha'  R^2+\beta_2\alpha^{\prime2}  R^3+\beta_3\alpha^{\prime3}  R^4
\right)
~,}
where $\kappa{}$ is the four-dimensional gravitational constant, and a Weyl transformation must be performed to bring the action to the 4d Einstein frame.\footnote{\label{f4}As emphasized in \cite{Grimm:2017okk}, 
in computing the 4d effective action the compactification must be performed around the  solution to the $\alpha'$-corrected equations of motion. 
This procedure can thus generate $\alpha'$-corrections also from the compactification of the ten-dimensional Einstein term.} 
Moreover each coefficient  in the series can be further expanded in the string coupling to separate the tree-level from the  one-loop contributions. 
Although all the higher-derivative terms in \eqref{scm} descend from the eight-derivative ten-dimensional $\alpha^{\prime3}$-corrections, they correspond to different orders of the 
4d  low-energy expansion. Indeed if $l_s=2\pi\sqrt{\alpha'}$,  $l_{4d}$ and $l_Y$ are the string length,  the four-dimensional low-energy  wavelength and the characteristic length of $Y$ respectively,  
we have,
\eq{\label{eqlim}
 {l_s^2}  \ll 
l^2_{Y}
\ll
l^2_{4d} 
~.}
Moreover the term with coefficient $\beta_n$ in \eqref{scm} is of order, 
\eq{\label{6}
\left(\frac{l_s}{l_{4d}} \right)^{2n}
\left(\frac{l_s}{l_{Y}}\right)^{6-2n}~;~n=1,2,3
~,}
relative to the Einstein term, 
so that the $n=1$ term dominates the $n=2,3$ terms in \eqref{scm}.

The ten-dimensional IIA supergravity (two-derivative) action admits solutions without flux of the form $\mathbb{R}^{1,3}\times Y$, where $Y$ is of $SU(3)$ holonomy (which for our purposes we take to be a compact CY). A sigma model argument \cite{Nemeschansky:1986yx} shows that this background can be promoted to  a solution to all orders in $\alpha'$, provided the metric of $Y$ is appropriately  corrected at each order  in such a way that it remains K\"{a}hler.\footnote{It should  be possible to generalize the sigma-model argument of  \cite{Nemeschansky:1986yx} to the case of backgrounds of the form $M_4\times Y$, where $M_4$ is an ALE space, along the lines of \cite{Bianchi:1994gi}.}  
Indeed \cite{Grimm:2017okk} confirms this to order $\alpha^{\prime3}$ and derives the explicit corrections to the dilaton and the metric, which is deformed away from Ricci-flatness at this order. Their derivation  remains valid for backgrounds of the form $M_4\times Y$, where 
$M_4$ is any Ricci-flat four-dimensional space.

Within the framework of  the effective 4d theory, nonperturbative gravitational instanton  corrections arise from vacua of the form $M_4\times Y$, where $M_4$ is an ALE space. 
These  instanton contributions are weighted by a factor $\exp(- S_0)$, where $S_0$ is the  4d effective action 
evaluated on  the solution $M_4\times Y$. Subject to the limitations discussed above, 
and taking into account the Ricci-flatness of the metric of $M_4$, 
the IIA 4d effective action of \cite{Grimm:2017okk}  reduces to,
\eq{\label{15}
2\kappa{}^2S_0=\beta_1\alpha' \int_{M_4}\d x^4\sqrt{g}  R_{\kappa\lambda\mu\nu}R^{\kappa\lambda\mu\nu} 
~,}
where in the conventions of \cite{Grimm:2017okk},\footnote{\label{f1}
The ten-dimensional gravitational constant of \cite{Grimm:2017okk} $2\kappa_{10}^2=(2\pi)^7\alpha^{\prime4}$, 
cf.~(2.4) therein,  is related to the four-dimensional one    
via $\kappa{}^2
={\kappa_{10}^2}/{l_s^6}$. Note in particular that 
eqs.~(4.9) and (4.19) of that reference are given in units where $l_s=2\pi\sqrt{\alpha'}=1$: to reinstate engineering dimensions 
one must  multiply with the appropriate powers of $l_s$. 
The 4d Einstein term  in \eqref{scm}  has been canonically normalized  via a Weyl transformation of the 4d metric. This 
affects the relative coefficient between two- and four-derivative terms in the action: note in particular that the right-hand side of \eqref{15} is invariant under Weyl transformations. 
We thank Kilian Mayer for clarifying to us 
the conventions of \cite{Grimm:2017okk}.}
\eq{\label{8bc}
\kappa{}^2
=\pi\alpha'~;~~~M_{\text{P}}=2\sqrt{\pi}~\!l_s^{-1}
~,}
with $M_{\text{P}}=\kappa^{-1}$ the (reduced) 4d Planck mass and 
$\beta_1$ given by,
\eq{\label{16}
l_s^6\beta_1=2^9\pi^4\alpha^{\prime2}\int_{Y}c_2\wedge J
~,}
where $c_2$ is the second Chern class of $Y$. 
For a generic K\"{a}hler manifold we have,
\eq{
c_2\wedge J=\frac{1}{32\pi^2}\left(
{R}_{mnkl}^2-\mathfrak{R}_{mn}^2+\frac14 \mathfrak{R}^2
\right)\text{vol}_6
~,}
where we have adopted real notation and defined $\mathfrak{R}_{mn}:={R}_{mnkl}J^{kl}$, $\mathfrak{R}:=\mathfrak{R}_{mn}J^{mn}$. 
The contractions are taken with respect to the metric compatible with  the K\"{a}hler form $J$ and the connection of the Riemann tensor.

The information about $Y$ enters the 4d effective action through the calculation of $\beta_1$. 
Since $\beta_1$ multiplies a term which is already a higher-order correction, it suffices to evaluate it in the CY limit (for which $\mathfrak{R}_{mn}$ vanishes). 
We thus obtain,
\eq{\label{b1}
\beta_1=\frac{1}{\pi^2l^2_s}\int_{Y}\d^6x  \sqrt{g}  ~\!{R}_{mnkl}^2>0
~.
}
Therefore the leading instanton contribution comes from the ALE space which minimizes the integral in \eqref{15}. 
This is the EH space \cite{Konishi:1989em}, cf.~\eqref{hirz}, 
so that,
\eq{\label{s0}
S_0=\frac{24}{\pi l_s^2}\int_{Y}\d^6x  \sqrt{g}  ~\!{R}_{mnkl}^2>0
~.}
Note that $S_0$ does not depend on the dilaton: this is related to the fact that, starting from an action of the form $\int\d^4 x\sqrt{g}(e^{-2\phi}R+\beta_1 \alpha' R_{\mu\nu\rho\sigma}^2)$, the dilaton exponential can be absorbed by a Weyl transformation of the form $g_{\mu\nu}\rightarrow e^{2\phi}g_{\mu\nu}$, cf.~footnote \ref{f1}. Therefore we have,
\eq{\label{s01}
S_0=c\left(\frac{l_Y}{l_{s}} \right)^{2}
~,}
with $c$ a positive number of order one.

\section{Consistent truncation}\label{sec:ct}

In \cite{Terrisse:2018qjm} we  presented a universal consistent truncation on Nearly-K\"{a}hler and CY manifolds in the presence of dilatino condensates. 
As it turns out, this consistent truncation captures only part of the universal scalar sector of the 
$\mathcal{N}=2$  low-energy effective supergravity obtained from IIA theory compactified on CY threefolds.  
Therefore we must  extend the ansatz of \cite{Terrisse:2018qjm} to include the ``missing'' fields and also to  take into account the gravitino condensates.

\subsection{Action and equations of motion}

In \cite{Soueres:2017lmy} the  quartic dilatino  terms of all (massive) IIA supergravities \cite{Giani:1984wc,Campbell:1984zc,Huq:1983im,Romans:1985tz,Howe:1998qt} were determined in the ten-dimensional superspace formalism of \cite{Tsimpis:2005vu}, and 
were found to agree with  \cite{Giani:1984wc}.  As follows from the result of \cite{Tsimpis:2005vu}, the quartic fermion terms are common to all IIA supergravities (massive or otherwise). In the following we will complete Romans supergravity (whose quartic fermion terms were not computed in \cite{Romans:1985tz}) by adding the quartic gravitino terms given in \cite{Giani:1984wc}.  
{}Furthermore we will set the dilatino to zero.  
Of course this would be inconsistent in general, since the dilatino couples linearly to gravitino terms. Here this does not lead to an inconsistency in the equations of motion, 
since we are ultimately interested in a maximally-symmettric vacuum, in which linear and cubic fermion VEV's vanish.

In the conventions of \cite{Soueres:2017lmy,Terrisse:2018qjm}, upon setting the dilatino to zero, the  action of Romans supergravity reads, 
\eq{\spl{\label{action3}
S=S_b&+\frac{1}{2\kappa_{10}^2}
\int\d^{10}x\sqrt{{g}} \Big\{ 
2(\tilde{\Psi}_M\Gamma^{MNP}\nabla_N\Psi_P)
+\frac{1}{2}e^{5\phi/4}m(\tilde{\Psi}_M\Gamma^{MN}\Psi_N) \\
&-\frac{1}{2\cdot 2!}e^{3\phi/4} F_{M_1M_2}(\tilde{\Psi}^M\Gamma_{[M}\Gamma^{M_1 M_2}\Gamma_{N]}\Gamma_{11}\Psi^N) \\
&-\frac{1}{2\cdot 3!}e^{-\phi/2} H_{M_1\dots M_3}(\tilde{\Psi}^M\Gamma_{[M}\Gamma^{M_1\dots M_3}\Gamma_{N]}\Gamma_{11}\Psi^N) \\
&+\frac{1}{2\cdot 4!}e^{\phi/4} G_{M_1\dots M_4}(\tilde{\Psi}^M\Gamma_{[M}\Gamma^{M_1\dots M_4}\Gamma_{N]}\Psi^N) 
+L_{\Psi^4}\Big\}
~,}}
where $\Psi_M$ is the gravitino; $S_b$ denotes the bosonic sector,  
\eq{\spl{\label{ba}S_b= \frac{1}{2\kappa_{10}^2}\int\d^{10}x\sqrt{{g}}\Big(
&-{R}+\frac12 (\partial\phi)^2+\frac{1}{2\cdot 2!}e^{3\phi/2}F^2\\
&+\frac{1}{2\cdot 3!}e^{-\phi}H^2+\frac{1}{2\cdot 4!}e^{\phi/2}G^2
+\frac{1}{2}m^2e^{5\phi/2}\Big) 
+\mathrm{CS}
~,
}}
and  CS is the Chern-Simons term. 
There are 24  quartic gravitino  terms as given in \cite{Giani:1984wc}, denoted $L_{\Psi^4}$ in \eqref{action3}. 
Of these only four can have a nonvanishing VEV in an ALE space: they are 
discussed in more detail in section \ref{sec:fermcond}.

We emphasize that the action \eqref{action3}  should 
be regarded as a 
book-keeping device whose variation with respect to the bosonic fields gives the correct bosonic
equations of motion in the presence of gravitino condensates. 
Furthermore, the
fermionic equations of motion are trivially satisfied in the maximally-symmetric vacuum. 
The  (bosonic) equations of motion (EOM) following from  (\ref{action3}) are as follows:

Dilaton EOM,
\eq{\spl{\label{beomf1}
0&=-{\nabla}^2\phi+\frac{3}{8}e^{3\phi/2}F^2-\frac{1}{12}e^{-\phi}H^2+\frac{1}{96}e^{\phi/2}G^2 +\frac{5}{4}m^2e^{5\phi/2}\\
&+\frac{5}{8}e^{5\phi/4}m(\tilde{\Psi}_M\Gamma^{MN}\Psi_N) \\
&-\frac{3}{16}e^{3\phi/4} F_{M_1M_2}(\tilde{\Psi}^M\Gamma_{[M}\Gamma^{M_1 M_2}\Gamma_{N]}\Gamma_{11}\Psi^N)\\
&+\frac{1}{24}e^{-\phi/2} H_{M_1\dots M_3}(\tilde{\Psi}^M\Gamma_{[M}\Gamma^{M_1\dots M_3}\Gamma_{N]}\Gamma_{11}\Psi^N) \\
&+\frac{1}{192}e^{\phi/4} G_{M_1\dots M_4}(\tilde{\Psi}^M\Gamma_{[M}\Gamma^{M_1\dots M_4}\Gamma_{N]}\Psi^N) 
~.
}}
Einstein EOM,
\eq{\spl{\label{beomf2}
{R}_{MN}&=\frac{1}{2}\partial_M\phi\partial_N\phi+\frac{1}{16}m^2e^{5\phi/2}{g}_{MN}
+\frac{1}{4}e^{3\phi/2}\Big(  2F^2_{MN} -\frac{1}{8} {g}_{MN}  F^2 \Big)\\
&+\frac{1}{12}e^{-\phi}\Big(  3H^2_{MN} -\frac{1}{4} {g}_{MN}  H^2 \Big)
+\frac{1}{48}e^{\phi/2}\Big(   4G^2_{MN} -\frac{3}{8} {g}_{MN}  G^2 \Big)\\
&+ \frac{1}{24}e^{\phi/4}G_{(M|}{}^{M_1M_2 M_3} 
 (\tilde{\Psi}_P\Gamma^{[P}\Gamma_{|N)M_1M_2 M_3}\Gamma^{Q]}\Psi_Q) \\
&-\frac{1}{96}e^{\phi/4}G_{M_1\dots M_4}\Big\{
(\tilde{\Psi}_P\Gamma_{(M}\Gamma^{M_1\dots M_4}\Gamma^{P}\Psi_{N)})-(\tilde{\Psi}_P\Gamma^P\Gamma^{M_1\dots M_4}\Gamma_{(M}\Psi_{N)})\\
&+\frac12 g_{MN}(\tilde{\Psi}^P\Gamma_{[P}\Gamma^{M_1\dots M_4}\Gamma_{Q]}\Psi^Q) 
\Big\}
-\frac18 g_{MN}L_{\Psi^4}+\frac{\delta L_{\Psi^4}}{\delta g^{MN}}
~,}}
where we have set: $\Phi^2_{MN}:=\Phi_{MM_2\dots M_p}\Phi_N{}^{M_2\dots M_p}$, for any $p$-form $\Phi$. 
In the Einstein equation above we have not included the gravitino couplings to the two- and three-forms: these vanish in the ALE background, as we will see in the following. 
Moreover, we have refrained from spelling out explicitly the quartic gravitino terms, as they are numerous and not particularly 
enlightening. We will calculate them explicitly later on in the case of the ALE space in section \ref{sec:fermcond}.

Form EOM's,\footnote{\label{f2}We are using 
``superspace conventions'' as in \cite{Lust:2004ig} so that,
\eq{
\Phi_{(p)}=\frac{1}{p!}\Phi_{m_1\dots m_p}\d x^{m_p}\swed\dots\swed\d x^{m_1}~;~~~
\d\Big( \Phi_{(p)} \swed\Psi_{(q)}\Big)=\Phi_{(p)} \swed\d\Psi_{(q)}
+(-1)^q\d\Phi_{(p)} \swed\Psi_{(q)}~.\nonumber
}
In $D$ dimensions the Hodge star is defined as follows,
\eq{
\star (\d x^{a_1}\wedge\dots\wedge \d x^{a_p})=\frac{1}{(D-p)!}\varepsilon^{a_1\dots a_p}{}_{b_1\dots b_{10-p}} \d x^{b_1}\wedge\dots\wedge \d x^{b_{10-p}}
~.\nonumber}
} 
\eq{\spl{\label{beomf3}
0&=\d {\star}\big[ e^{3\phi/2}F  
-\frac{1}{2}e^{3\phi/4} (\tilde{\Psi}^M\Gamma_{[M}\Gamma^{(2)}\Gamma_{N]}\Gamma_{11}\Psi^N)
\big]+ H\swed {\star} \big[e^{\phi/2}G {+ \frac{1}{2}e^{\phi/4}  (\tilde{\Psi}^M\Gamma_{[M}\Gamma^{(4)}\Gamma_{N]}\Psi^N)} \big]\\
0 &= \d{\star} \big[ e^{-\phi}H
-\frac{1}{2}e^{-\phi/2}  (\tilde{\Psi}^M\Gamma_{[M}\Gamma^{(3)}\Gamma_{N]}\Gamma_{11}\Psi^N)
\big]
+e^{\phi/2}F\swed {\star} \big[e^{\phi/2}G { + \frac{1}{2}e^{\phi/4}  (\tilde{\Psi}^M\Gamma_{[M}\Gamma^{(4)}\Gamma_{N]}\Psi^N)} \big]\\
 & -\frac{1}{2}G\swed G
+ m {\star}\big[ e^{3\phi/2}F  
{-\frac{1}{2}e^{3\phi/4} (\tilde{\Psi}^M\Gamma_{[M}\Gamma^{(2)}\Gamma_{N]}\Gamma_{11}\Psi^N)}
\big]\\
0&=\d
{\star} 
\big[
e^{\phi/2}G
   +\frac{1}{2}e^{\phi/4}  (\tilde{\Psi}^M\Gamma_{[M}\Gamma^{(4)}\Gamma_{N]}\Psi^N) \big]
-H\swed G
~,
}}
where
$\Gamma^{(p)}:=\frac{1}{p!}\Gamma_{M_1\dots M_p}\d x^{M_p}\wedge\dots\wedge\d x^{M_1}$. In addition the forms obey the  Bianchi identities,
\eq{\label{bi}
\d F= mH~;~~~\d H=0~;~~~\d G=H\wedge F
~.}

\subsection{Consistent truncation without condensates}\label{sec:consistent}

The truncation of \cite{Terrisse:2018qjm} contains the four real scalars $(A,\chi,\phi,\xi)$, with $A$ related to the volume modulus $v$ of section \ref{sec:review}: it  does not capture all the scalars of the 
universal sector of $\mathcal{N}=2$ supergravity, since it does not include the vectors and it truncates the two scalars $\xi'$, $b$ of section \ref{sec:review}. We must therefore expand the ansatz of \cite{Terrisse:2018qjm} to include the ``missing'' 
fields, at the same time taking the limit to the massless IIA theory, $m\rightarrow 0$. Explicitly we set,
\eq{\label{foranscy}
F=\d\alpha~;~~~ H=\d\chi \swed J+\d\beta~;~~~
G=\varphi\text{vol}_4+\frac12 c_0J\swed J+ J\swed (\d\gamma - \alpha\wedge \d\chi)-\frac{1}{2}\d\xi\swed\text{Im}\Omega
-\frac{1}{2}\d\xi'\swed\text{Re}\Omega
~,}
where $c_0$ is a real constant and $\varphi(x)$ is a 4d scalar. 
We have chosen to express $H$ in terms of the 4d potential $\beta$ instead of the axion. 
 Taking into account that for a  CY we have $\d J=\d\Omega=0$, this ansatz can be seen to automatically satisfy the Bianchi identities (\ref{bi}) in the massless limit. 
Our ansatz for the ten-dimensional metric reads,
\eq{\label{tdma}\d s^2_{(10)} =e^{2A(x)}\left(e^{2B(x)} g_{\mu\nu}\d x^{\mu}\d x^{\nu}+g_{mn}\d y^m\d y^n 
\right)~,
}
where the scalars $A$, $B$ only depend on the four-dimensional coordinates $x^\mu$. 
This gives,
\eq{\spl{
F^2_{\mu\nu} &= e^{-2A-2B} \d\alpha^2_{\mu\nu} ~;~~~ 
F^2 = e^{-4A-4B} \d\alpha^2 \\
H^2_{mn}&= 2e^{-4A-2B}(\partial\chi)^2g_{mn} ~;~~~
H^2_{\mu\nu} = 6e^{-4A}\partial_{\mu}\chi\partial_{\nu}\chi+e^{-4A-4B}h^2_{\mu\nu}\\
H^2 &= 18e^{-6A-2B}(\partial\chi)^2+e^{-6A-6B}h^2\\
G^2_{mn} &= 3e^{-6A-2B}\Big[(\partial\xi)^2+(\partial\xi')^2\Big]g_{mn}+12e^{-6A}c_0^2g_{mn}+3e^{-6A-4B}(\d\gamma - \alpha\wedge \d\chi)^2 g_{mn} \\
G^2_{\mu\nu} &=-6e^{-6A-6B}\varphi^2g_{\mu\nu}  + 6e^{-6A}( \partial_{\mu}\xi\partial_{\nu}\xi+\partial_{\mu}\xi'\partial_{\nu}\xi')+18e^{-6A-2B}(\d\gamma - \alpha\wedge \d\chi)^2_{\mu\nu}\\
G^2 &= -24 e^{-8A-8B}\varphi^2
+24e^{-8A-2B}\Big[(\partial\xi)^2+(\partial\xi')^2\Big]+72c_0^2e^{-8A}+36e^{-8A-4B}(\d\gamma - \alpha\wedge \d\chi)^2
~,}}
where the contractions on the left-hand sides above are computed with respect to the ten-dimensional metric; the contractions 
on the right-hand sides are taken with respect to the unwarped metric. It is also useful to note the following expressions,
\eq{\spl{\label{hodsr}
\star_{10} F &= \frac{1}{6}e^{6A} \star_4\d\alpha\swed J^3 
\\	
\star_{10} H &=  \tfrac12 e^{4A+2B} \star_{4}\!\d\chi\swed   J^2
+ \tfrac16 e^{4A-2B} \star_{4}\!h\swed   J^3
\\
\star_{10} G &= -\tfrac16 \varphi e^{2A-4B}  J^3
+c_0 e^{2A+4B} \text{vol}_4\swed J+\tfrac12 e^{2A}\star_{4}(\d\gamma - \alpha\wedge \d\chi)\swed J^2\\
&~~~~\!+\tfrac{1}{2} e^{2A+2B}   \star_{4}\!\d\xi\swed  \text{Re}\Omega
-\tfrac{1}{2} e^{2A+2B}   \star_{4}\!\d\xi'\swed  \text{Im}\Omega
~,}}
where the four-dimensional Hodge-star is taken with respect to the unwarped metric.

Plugging the above ansatz into the ten-dimensional EOM \eqref{beomf1}-\eqref{beomf3} we obtain the following: the internal $(m,n)$-components of the Einstein 
equations  read,
\eq{\spl{\label{et1}
0&=e^{-8A-2B}\nabla^{\mu}\left(
e^{8A+2B}\partial_{\mu}A
\right)
-\frac{1}{32} e^{3\phi/2-2A-2B} \d\alpha^2
+\frac18e^{-\phi-4A}(\partial\chi)^2
-\frac{1}{48}e^{-\phi-4A-4B}h^2\\
&-\frac{1}{32}e^{\phi/2-6A-2B}(\d\gamma - \alpha\wedge \d\chi)^2
+\frac{1}{16}e^{\phi/2-6A}\Big[(\partial\xi)^2
+(\partial\xi')^2\Big]\\
&+\frac{3}{16} 
e^{\phi/2-6A-6B}\varphi^2
+\frac{7}{16}e^{\phi/2-6A+2B}c_0^2
~.}}

The external $(\mu,\nu)$-components read,
\eq{\spl{\label{et2}
R^{(4)}_{\mu\nu}&=
g_{\mu\nu}\left(\nabla^{2}A+\nabla^{2} B+
8(\partial A)^2+2(\partial B)^2+10\partial A\cdot \partial B\right)
\\
&-8\partial_{\mu}A\partial_{\nu}A-2\partial_{\mu}B\partial_{\nu}B
-16\partial_{(\mu}A\partial_{\nu)}B+8\nabla_{\mu}\partial_{\nu}A+2\nabla_{\mu}\partial_{\nu}B\\
&+\frac32 e^{-\phi-4A} \partial_{\mu}\chi\partial_{\nu}\chi
+\frac12 e^{3\phi/2 -2A-2B} \d\alpha^2_{\mu\nu}
+\frac14 e^{\phi-4A-4B} h^2_{\mu\nu}
+\frac12 \partial_{\mu}\phi\partial_{\nu}\phi\\
&+\frac{1}{2} e^{\phi/2-6A}(\partial_{\mu}\xi\partial_{\nu}\xi+\partial_{\mu}\xi'\partial_{\nu}\xi')
+\frac{3}{2} e^{\phi/2-6A-2B}(\d\gamma - \alpha\wedge \d\chi)^2_{\mu\nu}
 \\
&+\frac{1}{16} g_{\mu\nu}\Big(  
- \frac{1}{2} e^{3\phi/2-2A-2B} \d\alpha^2
-\frac{1}{3}e^{\phi-4A-4B}h^2
-3e^{\phi/2-6A}\Big[(\partial\xi)^2
+(\partial\xi')^2
\Big]\\
&-6e^{-\phi-4A}(\partial\chi)^2
-5e^{\phi/2-6A-6B}\varphi^2 
-9c_0^2e^{\phi/2-6A+2B} -\frac{9}{2}e^{\phi/2-6A-2B}(\d\gamma - \alpha\wedge \d\chi)^2
\Big)
~,}}
while the mixed $(\mu,m)$-components are automatically satisfied. 
The dilaton equation reads,
\eq{\spl{\label{et3}
0&=e^{-10A-4B}\nabla^{\mu}\left(
e^{8A+2B}\partial_{\mu}\phi
\right)
-\frac{1}{4}e^{\phi/2-8A-2B}\Big[(\partial\xi)^2
+(\partial\xi')^2
\Big]
-\frac{3}{8}e^{3\phi/2-4A-4B} \d\alpha^2
\\
&+\frac32e^{-\phi-6A-2B}(\partial\chi)^2
+\frac{1}{12}e^{-\phi-6A-6B}h^2\\
&+\frac{1}{4}
e^{\phi/2-8A-8B}\varphi^2-\frac{3}{4}c_0^2 e^{\phi/2-8A}-\frac{3}{8}e^{\phi/2-8A-4B}(\d\gamma - \alpha\wedge \d\chi)^2
~.}}
The $F$-form equation of motion 
reduces to the condition,
\eq{
\d(e^{3\phi/2+6A} \star_4 \d\alpha) = \varphi e^{\phi/2+2A-4B}\d\beta - 3e^{\phi/2+2A}\d\chi\wedge\star_4(\d\gamma - \alpha\wedge \d\chi)
~.}
The $H$-form equation reduces to the following two equations,
\eq{\spl{\label{hfeom}
\d\left(
e^{-\phi+4A+2B}\star_4\d\chi
\right)  &= c_0\varphi \text{vol}_4 + (\d\gamma - \alpha\wedge \d\chi)\wedge(\d\gamma - \alpha\wedge \d\chi)-e^{\phi/2+2A} \d\alpha\wedge\star_4(\d\gamma - \alpha\wedge \d\chi)
~,}}
and, 
\eq{\label{seqh}
\d\left(
e^{-\phi+4A-2B}\star_4 \d\beta\right)
= 3c_0(\d\gamma - \alpha\wedge \d\chi)- \d\xi\wedge\d\xi' + e^{\phi/2+2A-4B} \varphi \d\alpha
~.}
The $G$-form equation of motion reduces to,
\eq{\spl{
\label{gfeom1}
\d\left(e^{\phi/2+2A+2B}\star_4\d\xi\right)  &=  h\swed\d\xi'\\
\d\left(e^{\phi/2+2A+2B}\star_4\d\xi'\right) &= -h\swed\d\xi\\
\d\left(e^{\phi/2+2A} \star_4(\d\gamma - \alpha\wedge \d\chi)\right) &=  2\d\chi\swed\d\gamma+c_0\d\beta
~,}}
together with the constraint,
\eq{\label{gfeom2}
0=\d\left(
 \varphi e^{\phi/2+2A-4B}+3c_0\chi
\right)
~.}
This can be readily integrated to give,
\eq{ \varphi= e^{-\phi/2-18A} (c_1-3c_0 \chi)~.}
Since $\chi$ only appears in the equations of motion through its derivatives or through $\varphi$, we may absorb $c_1$ by redefining $\chi$. This corresponds to a gauge transformation of the ten-dimensional $B$-field. We will thus set $c_1$ to zero in the following.

{\it The Lagrangian}

As we can see from (\ref{tdma}) the scalar $B(x)$ can be redefined away by absorbing it  in the 4d metric. This freedom can be exploited in order to 
obtain a 4d consistent truncation directly in the Einstein frame. The appropriate choice is, 
\eq{\label{baeq} B=-4A~.} 
With this choice one can  check that 
the ten-dimensional equations given in (\ref{et1})-(\ref{gfeom2}) all follow from the 4d  action,
%
%
\eq{\spl{
S_4=&\int\d^4 x\sqrt{g}
\Big(
R
 - 24 (\partial A)^2 
 -\tfrac{1}{2} (\partial \phi)^2  
-\tfrac{3}{2}  e^{-4A - \phi}(\partial \chi)^2 
- \tfrac{1}{2} e^{-6A + \phi/2} \left[(\partial \xi)^2+(\partial \xi')^2\right]\\
&-\tfrac{1}{4} e^{3\phi/2 + 6A} \d\alpha^2
-\tfrac{3}{4}  e^{\phi/2 + 2A} (\d\gamma - \alpha\swed\d\chi)^2
-\tfrac{1}{12} e^{-\phi + 12A} \d\beta^2
-\tfrac{9}{2}  e^{-\phi/2 - 18A} c_0^2 \chi^2
-\tfrac{3}{2}  e^{\phi/2 - 14A} c_0^2
 \Big)\\ 
&+\int 
3c_0 \d\gamma\swed\beta
+3c_0 \chi \ \alpha\swed\d\beta
+3 \chi\ \d\gamma\swed\d\gamma 
-\beta\swed\d\xi\swed\d\xi'
~.
}}
Furthermore equation \eqref{seqh} can be solved in order to express $\d\beta$ in terms of a scalar $b$ (the ``axion''),
\eq{\label{prhd}
 \d \beta =e^{\phi-12A}\star_4\left[\d b+\tfrac12(\xi\d\xi'-\xi'\d\xi) + 3c_0(\gamma - \chi \alpha)
\right]
~,}
where we chose the gauge most symmetric in $\xi$, $\xi'$. The Lagrangian becomes, in terms of the axion,
\eq{\spl{\label{36}
S_4=&\int\d^4 x\sqrt{g}
\Big(
R
 - 24 (\partial A)^2 
 -\tfrac{1}{2} (\partial \phi)^2  
-\tfrac{3}{2}  e^{-4A - \phi}(\partial \chi)^2 
- \tfrac{1}{2} e^{-6A + \phi/2} \left[(\partial \xi)^2+(\partial \xi')^2\right]\\
&-\tfrac{1}{4} e^{3\phi/2 + 6A} \d\alpha^2
-\tfrac{3}{4}  e^{\phi/2 + 2A} (\d\gamma - \alpha\swed\d\chi)^2
-\tfrac{1}{2} e^{\phi-12A} \left(\d b + \omega \right)^2\\
&-\tfrac{9}{2}  e^{-\phi/2 - 18A} c_0^2 \chi^2
-\tfrac{3}{2}  e^{\phi/2 - 14A} c_0^2
 \Big)
+\int 3\chi\ \d\gamma\swed\d\gamma 
~,
}}
where we have set,
\eq{
\omega:=\tfrac12(\xi\d\xi'-\xi'\d\xi) + 3c_0(\gamma - \chi \alpha)
~.} 
%

%

{\it Including background three-form flux}

We can include background three-form flux by modifying the form ansatz 
\eqref{foranscy} as follows,
\eq{\spl{\label{foranscyb}
F&=\d\alpha~;~~~ H=\d\chi \swed J+\d\beta+\frac12\text{Re}\big(b_0\Omega^*\big)\\
G&=\varphi\text{vol}_4+\frac12 c_0J\swed J+ J\swed (\d\gamma - \alpha\wedge \d\chi)-\frac{1}{2}D\xi\swed\text{Im}\Omega
-\frac{1}{2}D\xi'\swed\text{Re}\Omega
~,}}
where we have introduced a background charge $b_0\in\mathbb{C}$. The covariant derivatives are given by,
\eq{
D\xi:=\d\xi+b_1\alpha~;~~~D\xi':=\d\xi'+b_2\alpha
~,}
where we set $b_0=ib_1+b_2$. 
We see that the inclusion of a background charge for the three-form has the effect of gauging the isometries of the RR axions. 

The modified form ansatz \eqref{foranscyb} is such that it automatically satisfies the Bianchi identities. 
Moreover  the constraint \eqref{gfeom2} becomes,
\eq{\label{gfeom2mod}
0=\d\left(
 \varphi e^{\phi/2+18A}+3c_0\chi-\Xi
\right)
~,}
where we have set $\Xi:=b_2\xi-b_1\xi'$. As a consequence \eqref{prhd} gets modified,
\eq{\label{prhd2}
 \d \beta =e^{\phi-12A}\star_4\left[\d b+\tfrac12(\xi\d\xi'-\xi'\d\xi) + 3c_0(\gamma - \chi \alpha)+\Xi\alpha
\right]
~.}
The action reads,
\eq{\spl{\label{42}
S_4=&\int\d^4 x\sqrt{g}
\Big(
R
 - 24 (\partial A)^2 
 -\tfrac{1}{2} (\partial \phi)^2  
-\tfrac{3}{2}  e^{-4A - \phi}(\partial \chi)^2 
- \tfrac{1}{2} e^{-6A + \phi/2} \left[(D \xi)^2+(D \xi')^2\right]\\
&-\tfrac{1}{4} e^{3\phi/2 + 6A} \d\alpha^2
-\tfrac{3}{4}  e^{\phi/2 + 2A} (\d\gamma - \alpha\swed\d\chi)^2
-\tfrac{1}{2} e^{\phi-12A} \left(\d b + \tilde{\omega} \right)^2\\
&-\tfrac{1}{2}  e^{-\phi/2 - 18A} \big(3c_0 \chi -\Xi\big)^2
-\tfrac{1}{2}  e^{-\phi - 12A} |b_0|^2
-\tfrac{3}{2}  e^{\phi/2 - 14A} c_0^2
 \Big)
+\int 3\chi\ \d\gamma\swed\d\gamma 
~,
}}
where we have set,
\eq{\label{43}
\tilde{\omega}:=\tfrac12(\xi\d\xi'-\xi'\d\xi) + 3c_0(\gamma - \chi \alpha)+\Xi\alpha
~.} 

\subsection{Consistent truncation with condensates}\label{sec:fermcond}

In Euclidean signature the supersymmetric IIA action is constructed via the procedure of holomorphic complexification, see e.g.~\cite{Bergshoeff:2007cg}. 
This amounts to first expressing the Lorentzian action in terms of $\tilde{\Psi}_M$ instead of $\bar{\Psi}_M$ (which makes no difference in Lorentzian signature) and then Wick-rotating, 
see appendix \ref{app:spin} for our spinor and gamma-matrix conventions. In this way  one obtains a  (complexified) Euclidean action which is   formally identical to the Lorentzian one, with the difference  that now the two chiralities  ${\Psi}_M^{\pm}$,  should be thought of as independent complex spinors (there are no Majorana Weyl spinors in ten Euclidean dimensions). Although the gravitino ${\Psi}_M$ is complex in Euclidean signature, its complex conjugate does not appear in the action, hence the term  ``holomorphic complexification''.

Since we are interested in the case where only the 4d gravitino condenses, we expand the 10d gravitino as follows,
\eq{\label{grdc1}
\Psi_m=0~;~~~\Psi_{\mu+}=\psi_{\mu+}\otimes\eta-\psi_{\mu-}\otimes\eta^c~; ~~~\Psi_{\mu-}=\psi_{\mu+}'\otimes\eta^c-\psi_{\mu-}'\otimes\eta
~,}
so that,
\eq{
\tilde{\Psi}_{\mu+}=\tilde{\psi}_{\mu+}\otimes\tilde{\eta}+\tilde{\psi}_{\mu-}\otimes\tilde{\eta^c}~; ~~~
\tilde{\Psi}_{\mu-}=\tilde{\psi}_{\mu+}'\otimes\tilde{\eta^c}+\tilde{\psi}_{\mu-}'\otimes\tilde{\eta}
~.}
In Lorentzian signature the positive- and negative-chirality 4d vector-spinors above are related though complex conjugation: 
$\bar{\theta}_+^\mu=\tilde{\theta}_-^\mu$,  $\bar{\theta}_-^\mu=-\tilde{\theta}_+^\mu$, 
so that $\Psi_M$ is Majorana in 10d: $\bar{\Psi}_M=\tilde{\Psi}_M$. 
Upon Wick-rotating to Euclidean signature this is no longer true, and the two chiralities transform in independent representations. As already mentioned, in the present paper we focus on the contribution of ALE gravitational instantons to the fermion condensate. 
In this case there are no negative-chirality zeromodes and we can set, 
\eq{\label{posz}
\psi^{\mu}_-=\psi^{\prime\mu}_-=0~.
}
For any two 4d positive-chirality vector-spinors, $\theta_+^\mu$, $\chi_+^\mu$, the only nonvanishing bilinears read,
\eq{
\big(\theta^{[\mu_1}_+\gamma^{\mu_2\mu_3}\chi^{\mu_4]}_+\big)=\frac{i^s}{12} \varepsilon^{ \mu_1\mu_2\mu_3\mu_4 }
\big(\theta^{\lambda}_+\gamma_{\lambda\rho}\chi^{\rho}_+\big)~;~~~
\big(\theta^{\lambda}_+\chi_{\lambda+}\big)
~,
}
where  we used the Fierz identity \eqref{a4} and the Hodge duality relations \eqref{hm};   $s=1,2$ for Lorentzian, Euclidean signature respectively.  
Ultimately we will be interested in gamma-traceless vector-spinors, 
\eq{\label{zmgt}
\gamma_\mu\theta_+^\mu=\gamma_\mu\chi_+^\mu=0~,
}
since all ALE zeromodes can be put in this gauge \cite{Hawking:1979zs}. 
In this case we obtain the additional relation,
\eq{
\big(\theta^{\lambda}_+\gamma_{\lambda\rho}\chi^{\rho}_+\big)=
-\big(\theta^{\lambda}_+\chi_{\lambda+}\big)
~.}
Assuming, as is the case for ALE spaces, that only positive-chirality zeromodes exist in four dimensions, cf.~\eqref{posz}, 
the only nonvanishing bilinear condensates that appear in the equations of motion are proportional to,
\eq{\label{spbln}
\mathcal{A}
:=\left(\tilde{\psi}_{\mu+}\gamma^{\mu\nu}\psi'_{\nu+}\right)
=-\left(\tilde{\psi}^{\mu}_+\psi'_{\mu+}\right)
~,}
where in the second equality we have assumed that $\psi^\mu_{+}$, $\psi^{\prime \mu}_{+}$ are gamma-traceless, cf.~\eqref{zmgt}. 

Furthermore we note   
the following useful results,
\eq{\spl{\label{grbilfg}
\left(\tilde{\Psi}_\rho\Gamma_{(\mu}\Gamma^{M_1\dots M_4}\Gamma^{\rho}\Psi_{\nu)}\right)G_{M_1\dots M_4}&=24(3c_0e^{-4A}+\varphi e^{-4A-4B})\mathcal{A}g_{\mu\nu}\\
\left(\tilde{\Psi}_\rho\Gamma_{\sigma}\Gamma_{(\mu}{}^{M_2M_3 M_4}\Gamma^{\rho}\Psi^{\sigma}\right)G_{\nu)M_2M_3 M_4}&=24 \varphi e^{-4A-4B}\mathcal{A}g_{\mu\nu}\\
\left(\tilde{\Psi}_\rho\Gamma_{\sigma}\Gamma_{(m}{}^{M_2M_3 M_4}\Gamma^{\rho}\Psi^{\sigma}\right)G_{n)M_2M_3 M_4}&=48 c_0\mathcal{A}e^{-4A-2B}g_{mn}
~,}}
where on the left-hand sides above we used the warped metric for the contractions, while on the right-hand sides we used the unwarped metric. 
In the 4d theory, these bilinears receive contributions from the EH instanton at one loop in the gravitational coupling.

In the presence of gravitino condensates the equations of motion \eqref{et1}-\eqref{gfeom2} are modified as follows:
 the internal $(m,n)$-components of the Einstein 
equations  read,
\eq{\spl{\label{et12}
0&=e^{-8A-2B}\nabla^{\mu}\left(
e^{8A+2B}\partial_{\mu}A
\right)
+\dots+\frac{1}{4}\left( \varphi  e^{\phi/4-4A-4B}-c_0e^{\phi/4-4A}
  \right)\mathcal{A}-\frac18 e^{2A+2B}L_{\Psi^4}
~,}}
where the ellipses stand for terms that are identical to the case without fermion condensates. 
The external $(\mu,\nu)$-components read,
\eq{\spl{\label{et22}
R^{(4)}_{\mu\nu}&=
\dots
-\frac{1}{2} g_{\mu\nu} e^{\phi/4-4A-4B} \varphi \mathcal{A}
~,}}
while the mixed $(\mu,m)$-components are automatically satisfied. 
The dilaton equation reads,
\eq{\spl{\label{et32}
0&=e^{-10A-4B}\nabla^{\mu}\left(
e^{8A+2B}\partial_{\mu}\phi
\right)+\dots 
+\tfrac{1}{4} (3c_0e^{\phi/4+2A}+\varphi e^{\phi/4+2A-4B})\mathcal{A}
~.}}
%
%
The $F$-form and $H$-form equations are modified as follows,
\eq{\label{ffeom2}
\d(e^{3\phi/2+6A} \star_4 \d\alpha) = \dots  {+ e^{\phi/4+4A-2B} \mathcal{A}\ \d\beta}
~,}
and,
\eq{\label{seqh2}
\d\left(
e^{-\phi+4A-2B}\star_4 \d\beta\right) = \dots  { + e^{\phi/4+4A-2B} \mathcal{A}\ \d\alpha}
~,}
respecively.  The $G$-form equation of motion remains unchanged except for the constraint,
\eq{\label{gfeom22}
0=\d\left(
 \varphi e^{\phi/2+2A-4B}
+ 3c_0 \chi-\Xi
+ e^{\phi/4+4A-2B}  \mathcal{A}
\right)
~.}
In deriving the above we have taken into account that,
\eq{
(\tilde{\Psi}^M\Gamma_{[M}\Gamma^{(4)}\Gamma_{N]}\Psi^N) =
2\mathcal{A} e^{2A+2B}\left(
\text{vol}_4-\frac12 e^{-4B}J\swed J 
\right)
~.}
At this stage it is important to notice that the new $\mathcal{A}$ terms in the flux equations (\ref{ffeom2}) and (\ref{seqh2}) exactly compensate the modification of $\varphi$ in  (\ref{gfeom22}), so that the form equations are ultimately unchanged in the presence of fermion condensates.

Of the 24 quartic gravitino terms that appear in the action of \cite{Giani:1984wc} only the following are nonvanishing,
\eq{\spl{\label{grpernici}
\left(\tilde{\Psi}_\mu\Gamma_{11}\Psi_{\nu}\right) \left(\tilde{\Psi}^\mu\Gamma_{11}\Psi^{\nu}\right)&=
4\left(\tilde{\psi}^{[\mu}_{+} \psi^{\prime \nu]}_{+}\right)^2e^{-4A-4B}\\
\left(\tilde{\Psi}^{\mu_1}\Gamma_{11}\Gamma_{\mu_1\dots\mu_4}\Psi^{\mu_2}\right) \left(\tilde{\Psi}^{\mu_3}\Gamma_{11}\Psi^{\mu_4}\right)&=
-\frac16\left(\tilde{\Psi}^{\mu_1}\Gamma_{\mu_1\dots\mu_4mn}\Psi^{\mu_2}\right) \left(\tilde{\Psi}^{\mu_3}\Gamma^{mn}\Psi^{\mu_4}\right)\\
&=
-\left(8\tilde{\psi}_{[\nu+} \psi'_{\rho]+}+4\tilde{\psi}^{\mu}_+ \gamma_{\rho\nu}\psi'_{\mu+}\right)
\left(\tilde{\psi}^{\rho}_+ \psi_+^{\prime \nu}\right)e^{-4A-4B}\\
\left(\tilde{\Psi}^{[M_1}\Gamma^{M_2M_3}\Psi^{M_4]}\right)^2  &=
4\left(\tilde{\psi}_+^{[\mu_1}\gamma^{\mu_2\mu_3}\psi_+^{\prime \mu_4]}\right)^2e^{-4A-4B}
-\frac23 \left(\tilde{\psi}^{[\mu}_{+} \psi^{\prime \nu]}_{+}\right)^2e^{-4A-4B}~,}}
where for the contractions on the left-, right-hand sides above we have used the warped, unwarped metric respectively. 
We thus obtain, cf. \eqref{action3},
\eq{\spl{\label{qred1}
L_{\Psi^4} &=
 \frac{1}{4} (\tilde{\Psi}_M \Gamma_{11} \Psi_N)^2
+\frac{1}{8} \tilde{\Psi}^{M_1}\Gamma_{11} \Gamma_{M_1 \cdots M_4} \Psi^{M_2} \ \tilde{\Psi}^{M_3}\Gamma_{11} \Psi^{M_4}\\
&+\frac{1}{16} \tilde{\Psi}^{M_1} \Gamma_{M_1\cdots M_6} \Psi^{M_2} \ \tilde{\Psi}^{M_3} \Gamma^{M_4M_5} \Psi^{M_6}
+\frac{3}{4} (\tilde{\Psi}_{[M_1} \Gamma_{M_2M_3} \Psi_{M_4]})^2\\
           &= e^{-4A-4B} \mathcal{B}
           ~,}}
where we have defined,
\eq{\label{calbdef}
\mathcal{B}:= 
 -\frac{3}{2} (\tilde{\psi}_{[\mu}\psi'_{\nu]})^2
 + (\tilde{\psi}^\mu \gamma_{\rho\nu} \psi'_\mu) (\tilde{\psi}^\rho \psi'^\nu) 
 +3 (\tilde{\psi}_{[\mu_1} \gamma_{\mu_2\mu_3}\psi'_{\mu_4]})^2
~,}
which does not depend on the warp factor. 
In the 4d theory,  at one-loop order in the gravitational coupling, 
the quartic gravitino term receives contributions  from the ALE instanton with $\tau=2$ (four spin-3/2 zeromodes).

{\it The Lagrangian}

Imposing \eqref{baeq} as before, and solving once again for $\varphi$,
\eq{\label{wnth}\varphi = e^{-\phi/2-18A} \left(\Xi -3c_0\chi  -e^{\phi/4+12A}\mathcal{A}  \right) ~,}
where $\Xi$ was defined below \eqref{gfeom2mod}, 
it can now be seen that the ten-dimensional equations in the presence of gravitino condensates all follow from the 4d action,
%
%
\eq{\spl{
S_4=\int\d^4 x\sqrt{g}&
\Big(
R
 - 24 (\partial A)^2 
 -\tfrac{1}{2} (\partial \phi)^2  
-\tfrac{3}{2}  e^{-4A - \phi}(\partial \chi)^2 
- \tfrac{1}{2} e^{-6A + \phi/2} \left[(D \xi)^2+(D \xi')^2\right]\\
&-\tfrac{1}{4} e^{3\phi/2 + 6A} \d\alpha^2
-\tfrac{3}{4}  e^{\phi/2 + 2A} (\d\gamma - \alpha\swed\d\chi)^2
-\tfrac{1}{12} e^{-\phi + 12A} \d\beta^2-V
 \Big)\\ 
+\int 
&3c_0 \d(\gamma-\alpha\chi)\swed\beta
+3 \chi\ \d\gamma\swed\d\gamma 
+\Xi\beta\swed\d\alpha-\beta\swed D\xi\swed D\xi'
~,
}}
where the potential of the theory is given by,
\boxedeq{\spl{\label{58}
V(\chi,\xi,\xi',\phi,A)
=
&\tfrac{3}{2}  c_0^2e^{\phi/2 - 14A}
 + \tfrac{1}{2}|b_0|^2e^{-\phi-12A}
-3c_0 {\mathcal{A}}e^{\phi/4 - 4A}
+e^{6A}\mathcal{B} \\
+&\tfrac12\Big(
{\mathcal{A}}e^{3A}+(3c_0\chi -\Xi)e^{-\phi/4 - 9A}
\Big)^2
~.
}}
Note that in integrating the 4d Einstein equation \eqref{et22}, care must be taken  to first substitute in the right-hand side the value of $\varphi$ from \eqref{wnth}, and 
take into account the variation of the condensates $\mathcal{A}$, $\mathcal{B}$ with respect to the metric.

The modifications due to the condensate in \eqref{seqh2} and \eqref{wnth} are such that the relation (\ref{prhd2}) between $\beta$ and the axion is unchanged. In terms of the axion, the action reads,
\boxedeq{\spl{\label{ctr2}
S_4=\int&\d^4 x\sqrt{g}
\Big(
R
 - 24 (\partial A)^2 
 -\tfrac{1}{2} (\partial \phi)^2  
-\tfrac{3}{2}  e^{-4A - \phi}(\partial \chi)^2 
- \tfrac{1}{2} e^{-6A + \phi/2} \left[(D \xi)^2+(D \xi')^2\right]\\
&-\tfrac{1}{4} e^{3\phi/2 + 6A} \d\alpha^2
-\tfrac{3}{4}  e^{\phi/2 + 2A} (\d\gamma - \alpha\swed\d\chi)^2
-\tfrac{1}{2} e^{\phi - 12A} (\d b + \tilde{\omega})^2
-V
 \Big) 
+\int 
3 \chi\ \d\gamma\swed\d\gamma  
~,
}}
where $\tilde{\omega}$ was defined in \eqref{43}. Note that $\chi$, $\xi$, $\xi'$ enter the potential  only through the linear combination $3c_0\chi -\Xi$, so 
two of these scalars remain flat directions even in the presence of the flux and the condensate, just as the  axion $b$.

\subsection{Vacua}\label{sec:vacua}

Maximally-symmetric solutions of the effective 4d theory \eqref{ctr2} can be obtained by setting the vectors to zero,
\eq{\alpha=\gamma=0~,}
and minimizing the potential of the theory,
\eq{\label{mincond}
\overrightarrow{\nabla} V(\chi_0,\xi_0,\xi'_0,\phi_0,A_0)=0~,
}
where $(\chi,\xi,\xi',\phi,A)=(\chi_0,\xi_0,\xi'_0,\phi_0,A_0)$  is the location of the minimum in field space. Then the Einstein equations 
determine the scalar curvature of the 4d spacetime  to be,\footnote{Note that \eqref{mcD} is different from the standard relation $R=2V_0$. This is because the condensates $\mathcal{A}$, $\mathcal{B}$ have non-trivial variations with respect to the metric.}
\eq{\spl{\label{mcD}
R&=  {3}  c_0^2e^{\phi_0/2 - 14A_0}
 + |b_0|^2e^{-\phi_0-12A_0}
-3c_0 {\mathcal{A}}e^{\phi_0/4 - 4A_0}\\
&+( 3c_0\chi_0-\Xi_0)^2e^{-\phi_0/2 - 18A_0}+(3c_0\chi_0-\Xi_0) {\mathcal{A}}e^{-\phi_0/4 - 6A_0}
~,}}
where $\Xi_0=b_2\xi_0-b_1\xi'_0$, and  we assume that a Wick rotation has been performed back to Minkowski signature. 

Condition \eqref{mincond} admits two classes of solutions.


{\it Case 1}:  $c_0=0$

In this case imposing \eqref{mincond} sets $b_0=0$, and 
the potential only depends on the warp factor $A$. 
A minimum is obtained at finite value of $A$ provided,
\eq{
{\mathcal{B}}= -\tfrac12 {\mathcal{A}}^2 
~,}
and requires the quartic condensate to be negative. From  \eqref{mcD} it then follows that $R=0$, and we obtain a Minkowski 4d vacuum. In fact the potential vanishes identically.

{\it Case 2}:  $c_0\neq0$

In this case  \eqref{mincond} can be solved for finite values of $\phi$ and $A$. 
The value of $\chi$ at the minimum is given by,
\eq{\chi_0=- \frac{1}{3c_0} ~\!\left( \mathcal{A}g_s^{1/4}e^{ 12A_0}
-\Xi_0
\right)
~,}
where we have set $g_s:=e^{\phi_0}$. Minimization of $V$ with respect to $\xi$, $\xi'$ does not give additional constraints, so that $\xi_0$, $\xi'_0$ remain undetermined. 
The values of $\phi_0$ and $A_0$ at the minimum can also  be adjusted arbitrarily, and determine $|b_0|$ and $c_0$ in terms of the condensates,
\eq{\spl{\label{63a}
|b_0|^2 &=
\frac{3}{400}~\! g_se^{18A_0}\left( 
 40 \mathcal{B} - 21 \mathcal{A}^2 \mp 3\mathcal{A} \sqrt{ 49\mathcal{A}^2+80\mathcal{B} }
 \right)\\
c_0     &=
\frac{1}{20}~\!g_s^{-1/4}e^{10A_0}\left(
7\mathcal{A}\pm\sqrt{ 49\mathcal{A}^2+80\mathcal{B} }
\right)
~,}}
where  the signs in $b_0$ and $c_0$ are correlated. 
Henceforth we will set $e^{A_0}=1$, since the warp factor at the minimum can be absorbed in $l_Y$.

Consistency of \eqref{63a} requires the quartic condensate to obey the constraint, 
\eq{\label{65r}
\mathcal{B}> 0
~,}
and correlates the sign of $\mathcal{A}$  with the two branches of the solution: 
the upper/lower  sign  in \eqref{63a} corresponds to $\mathcal{A}$ negative/positive, respectively.\footnote{If $\mathcal{B}> 3\mathcal{A}^2/2$, we may also take  the upper/lower  sign  in \eqref{63a} for $\mathcal{A}$ positive/negative, respectively.  Equation \eqref{65r} is the  weakest condition on the quartic condensate that is sufficient for consistency of the solution.}

From \eqref{mcD} it then follows that,
\eq{\label{70ad}
R_{\text{dS}}=  3 g_s^{-1} |b_0|^2
\propto l_s^{-2}e^{-2c~\!(l_Y/l_s)^2}
~,
} 
up to a proportionality constant of order one. 
We thus obtain a de Sitter 4d vacuum, provided \eqref{65r} holds. 
In the equation above we have taken into account that the quadratic and quartic condensates are 
expected to be of the general  form, cf.~the discussion around \eqref{68},
\eq{\label{abv}
 {\mathcal{A}}\propto l_s^{-1}e^{-c~\!(l_Y/l_s)^2}~;~~~  {\mathcal{B}}\propto l_s^{-2}e^{-2c~\!(l_Y/l_s)^2}
~,
}
up to proportionality constants of order one.

We have verified numerically, as a function of $\mathcal{A}^2/\mathcal{B}$,  that
all three eigenvalues of 
 the Hessian of the potential are positive at the solution. 
 I.e. the solution is a local minimum of the potential \eqref{58}.

{\it Flux quantization}

The four-form flux is constrained to obey,\footnote{The Page form corresponding to $G$ is given by $\hat{G}:=G-H\wedge \alpha$, which is closed.  The difference between  
$G$ and $\hat{G}$ vanishes when integrated over four-cycles of $Y$.}
\eq{
\frac{1}{l_s^3}\int_{\mathcal{C}_A}G\in\mathbb{Z}
~,}
where $\{ \mathcal{C}_A~;~A=1,\dots,h^{2,2}\}$  is a basis of integral four-cycles of the CY, $\mathcal{C}_A\in H_4(Y,\mathbb{Z})$. From \eqref{foranscyb}, \eqref{63a}, \eqref{abv} we then obtain,
\eq{\label{67b}
n_A \propto  g_s^{-1/4}\Big(\frac{l_Y}{l_s}\Big)^4  e^{-c~\!(l_Y/l_s)^2}\text{vol}(\mathcal{C}_A)
~,}
up to a proportionality constant of order one; 
$\text{vol}(\mathcal{C}_A)$ is the volume of the four cycle $\mathcal{C}_A$ in units of $l_Y$, and $n_A\in\mathbb{Z}$. 
Since the string coupling can be tuned to obey  $g_s\ll 1$ independently of the $l_Y/l_s$ ratio, 
\eqref{67b} can be solved for  $\text{vol}(\mathcal{C}_A)$ of order one, provided we take $n_A$ sufficiently close to each other. Given a set of flux quanta $n_A$, this equation 
fixes the K\"{a}hler moduli in units of $l_Y$; the overall CY volume is set by $l_Y$, which remains unconstrained.

Note that  even if we allow for large flux quanta in order to solve the flux quantization constraint, it can be seen that higher-order flux corrections are subdominant 
in the $g_s\ll1$ limit. Indeed the parameter that controls the size of these corrections is $|g_sG|$, which scales as $g_s^{3/4}$.

Similarly, the three-form flux is constrained to obey,
\eq{
\frac{1}{l_s^2}\int_{\mathcal{C}_\alpha}H\in\mathbb{Z}
~,}
where $\{ \mathcal{C}_\alpha~;~A=1,\dots,h^{2,1}\}$  is a basis of integral three-cycles of the CY, $\mathcal{C}_\alpha\in H_3(Y,\mathbb{Z})$. From \eqref{foranscyb} we can see 
that this equation constrains the periods of $\Omega$, and hence the complex-structure moduli of $Y$.

\section{Discussion}\label{sec:discussion}

We considered the effect of gravitino condensates from ALE instantons, in the context of a 4d consistent truncation of IIA on CY in the presence of background flux. 
The 4d theory admits de Sitter solutions, which are local minima of the potential \eqref{58}, provided the quartic condensate  has a positive sign, cf.~\eqref{65r}. We do not know whether or not this is the case, as this would require  knowledge of 
the explicit form of the zero modes of the Dirac operator in the  $\tau=2$ ALE background. Clearly it would be crucial to construct these zero modes (which, to are knowledge, have never been explicitly computed), generalizing 
the calculations of 
\cite{Hawking:1979zs,Konishi:1988mb,Bianchi:1994gi} 
to the second gravitational instanton in the ALE series.

The validity of the de Sitter solutions presented here requires the higher-order string-loop corrections in the 4d action to be subdominant with respect to the ALE instanton contributions to the gravitino condensates.  Since the latter do not depend on  the string coupling, cf.~\eqref{abv}, there is no obstruction to tuning $g_s$ to be sufficiently small, 
$g_s\ll 1$, in order 
for the string-loop corrections to be negligible with respect to the instanton contributions.

The $l_Y/l_s$ ratio can be tuned so that the condensates are of the order of the Einstein term in the 4d action, thus dominating 4d higher-order derivative corrections. This requires,
\eq{\label{75tyu}
l_{4d}^{-2}\sim R_{\text{dS}}
\propto
l_s^{-2}e^{-2c~\!(l_Y/l_s)^2}
~,}
where we have taken \eqref{70ad} into account. Current cosmological data give, 
\eq{
 \frac{ R_{\text{dS}} }{M_{\text{P}}^2 } 
\sim\Big( \frac{l_s}{l_{4d} } \Big)^2\sim10^{-122}
~.
}
From \eqref{75tyu} we then obtain $l_Y/l_s\sim 10$ for  $c$ of order one, cf.~\eqref{s01}.

In addition to the higher-order derivative corrections, the 4d effective action receives corrections at the two-derivative level, of the form $(l_s/l_Y)^{2n}$ with $n\geq1$. These come from 
a certain subset of the 10d tree-level 
$\alpha'$ corrections (string loops are subleading), which include the $R^2(\partial F)^2$ corrections of  \cite{Policastro:2006vt}. Given the $l_Y/l_s$ ratio derived above, these corrections will be of 
the order of one percent or less.

As is well known, the vacua computed within the framework of consistent truncations, such as the one constructed in the present paper, are susceptible to 
destabilization by modes that are truncated out of the spectrum. This is an issue that needs to be addressed before one  can be confident 
of the validity of the vacua presented here. The stability issue is particularly important given the fact that, in the presence of a non-vanishing 
 gravitino condensate, supersymmetry will generally be broken.

Ultimately,  the scope of  the path integral over metrics approach to quantum gravity is limited, since  the 4d gravity theory is  non-renormalizable.  Rather  it should be thought of as an effective low-energy limit of string theory. A natural approach to gravitino condensation from the string/M-theory standpoint, would be to try to construct brane-instanton analogues of the four-dimensional gravitational instantons. The fermion condensates might then be computed along the lines of \cite{Becker:1995kb,Harvey:1999as,Tsimpis:2007sx}.

Another interesting direction would be to try to embed the consistent truncation of the present paper within the framework 
of $\mathcal{N}=2$ 4d (gauged) supergravity.  
On general grounds \cite{Cvetic:2000dm}, we expect the existence of a consistent truncation of a higher-dimensional
supersymmetric theory to the bosonic sector of a supersymmetric lower-dimensional
theory, to guarantee the existence of a consistent truncation to the full lower-dimensional theory. 
The condensate would then presumably be associated with certain gaugings of the 4d theory.

\section*{Acknowledgment}

We would like to thank  Thomas Grimm and  Kilian Mayer for useful correspondence.

\appendix

\section{Conventions}\label{app:spin}

As in \cite{Terrisse:2018qjm}, our spinor conventions are  those listed in appendix A of \cite{Lust:2004ig}, except that the 
$SU(3)$-singlet spinor of the internal manifold is denoted $\eta$ here and corresponds to the  $\eta_+$  of \cite{Lust:2004ig}. 
Moreover $\eta_-$ of \cite{Lust:2004ig} corresponds to $\eta^c:=C\eta^*$ here.

Our conventions for the explicit 4d spinor indices  are as follows. A positive-, negative-chirality 4d Weyl spinor is indicated with a lower, upper spinor index respectively: 
$\theta_\alpha$, $\chi^\alpha$. We never raise or lower the spinor indices on spinors, so that the position unambiguously indicates the chirality. 
The 4d gamma matrices, the  charge conjugation and chirality matrices are decomposed into chiral blocks,
\eq{\label{a1}
\gamma_\mu=
\left( {\begin{array}{cc}
   0 & \left(\gamma_\mu\right)_{\alpha\beta} \\
   \left(\gamma_\mu\right)^{\alpha\beta} & 0
  \end{array} } \right)~;~~~
  C^{-1}=
  \left( {\begin{array}{cc}
  C^{\alpha\beta} & 0 \\
  0 & C_{\alpha\beta}
  \end{array} } \right)~;~~~
    \gamma_5=
  \left( {\begin{array}{cc}
  \delta_{\alpha}{}^{\beta} & 0 \\
  0 & -\delta^{\alpha}{}_{\beta}
  \end{array} } \right)
~.}
It is the ``Pauli matrices'' $(C^{-1}\gamma_{\mu_1\dots\mu_n})$ which act as Clebsch-Gordan coefficients between spinor bilinears and $n$-forms. 
For example, the structure of indices of the charge conjugation matrix reflects the fact that scalars can only be formed as 
spinor bilinears of Weyl spinors of the same chirality,
\eq{\label{a2}
v=  \theta_\alpha  C^{\alpha\beta} \chi_\beta~;~~~
u=  \theta^\alpha  C_{\alpha\beta} \chi^\beta
~.}
As another example, the structure of indices of $C^{-1}\gamma_\mu$ reflects the fact that vectors can only be formed as spinor bilinears of  Weyl spinors of opposite chirality,
\eq{\label{a3}
v_\mu=  \theta^\alpha  \left(C^{-1}\gamma_\mu\right)_\alpha{}^{\beta}\chi_\beta~;~~~
u_\mu=  \theta_\alpha  \left(C^{-1}\gamma_\mu\right)^\alpha{}_{\beta}\chi^\beta
~.}
We also make use of the Fierz relation for two positive-chirality 4d spinors,
\eq{\label{a4}
\theta_\alpha\chi_\beta=-\frac12(\tilde{\theta}\chi)C_{\alpha\beta}-\frac18(\tilde{\theta}\gamma_{\mu\nu}\chi)\left(\gamma^{\mu\nu}C\right)_{\alpha\beta}
~,}
where $\tilde{\theta}\equiv \theta^{\text{Tr}}C^{-1}$, and similarly for negative chirality.

The Hodge duality relations read,
\eq{\label{hm}
\tfrac{1}{(4-l)!}~\!\varepsilon_{\mu_1\dots\mu_{l}}{}^{\nu_1\dots\nu_{4-l}}\gamma_{\nu_1\dots\nu_{4-l}}=i^s(-1)^{\frac12l(l-1)}\gamma_{\mu_1\dots\mu_{l}}\gamma_5
~,}
where $s=1,2$ for Lorentzian, Euclidean signature respectively.
With explicit spinor indices in Euclidean signature we have,
\eq{\label{hodge}
\frac12\varepsilon_{\mu\nu\rho\sigma}\left(\gamma^{\rho\sigma}\right)_\alpha{}^\beta=\left(\gamma_{\mu\nu}\right)_\alpha{}^\beta~;~~~
\frac12\varepsilon_{\mu\nu\rho\sigma}\left(\gamma^{\rho\sigma}\right)^\alpha{}_\beta=-\left(\gamma_{\mu\nu}\right)^\alpha{}_\beta
~.}
In particular if $T_{\mu\nu}$ is a self-dual tensor, $\tfrac12\varepsilon_{\mu\nu\rho\sigma}T^{\rho\sigma}=T_{\mu\nu}$, 
it follows that $T\cdot\gamma$ vanishes when acting on negative-chirality spinors,
\eq{\label{sdi}
T^{\mu\nu}\left(\gamma_{\mu\nu}\right)^\alpha{}_\beta=0
~.}

\section{ALE instantons}\label{app:ale}

Asymptotically locally Euclidean (ALE) spaces, see e.g.~\cite{Eguchi:1980jx} for a review,  are noncompact self-dual gravitational instantons, i.e.~their Riemann tensor obeys,
\eq{\label{51}
\frac12\varepsilon_{\mu\nu\rho\sigma}R_{\kappa\lambda}{}^{\rho\sigma}=R_{\kappa\lambda\mu\nu}
~.}
From the above and the identity $R_{[\kappa\lambda\mu]\nu}=0$, it follows that the ALE spaces are Ricci-flat,
\eq{\label{rf}
R_{\mu\nu}=0
~.}
These spaces asymptote $S^3/\mathbb{Z}_{k+1}$ at spatial infinity, with $k\in\mathbb{N}$ (the case $k=0$ corresponds to $\mathbb{R}^4$).  
The simplest nontrivial example in this class is the EH space \cite{Eguchi:1978gw}, which corresponds to $k=1$.  
Explicitly the metric reads,
\eq{\label{m}
\d s^2=
\d r^2\big(1-\tfrac{a^4}{r^4}\big)^{-1}
+\tfrac14 r^2\Big(  
\sigma_1^2+\sigma_2^2+
\big(1-\tfrac{a^4}{r^4}\big)\sigma_3^2
\Big)
~,}
where $a>0$ is an arbitrary constant, and,
\eq{\label{52}
\sigma_1=\sin\psi\d\theta-\sin\theta\cos\psi\d\phi~;~~~
\sigma_2=-\cos\psi\d\theta-\sin\theta\sin\psi\d\phi~;~~~
\sigma_3=\d\psi+\cos\theta\d\phi~.
}
For  the coordinate ranges $a \leq r$,  $0\leq\theta\leq\pi$, $0\leq\phi\leq2\pi$, $0\leq\psi\leq2\pi$, the manifold can be seen to be smooth  with boundary given by 
$\mathbb{RP}^3=S^3/\mathbb{Z}_2$ at asymptotic infinity. (We would have an asymptotic $S^3$ if $0\leq\psi\leq4\pi$). The Hirzebruch signature $\tau$ of a self-dual space is given by, 
\eq{\label{hirz}
\tau= \frac{1}{48\pi^2}\int\d x^4\sqrt{g}R_{\kappa\lambda\mu\nu}R^{\kappa\lambda\mu\nu} \in\mathbb{N}~.}
As can be verified using \eqref{m},  the EH  gravitational instanton is the ALE space with the smallest Hirzebruch signature, $\tau=1$. 
More generally it can be shown that $\tau=k$, with $k$ as given below eq.~\eqref{rf}.

It is convenient to use a gauge in which not only the curvature but also the connection is self-dual \cite{Eguchi:1980jx},
\eq{\label{sdom}
\omega_{ab}=\frac12\varepsilon_{abcd}\omega^{cd}
~.}
In this gauge the covariant derivative reduces to a simple derivative on negative chirality spinors,
\eq{
\nabla_\mu\theta^\alpha=\partial_\mu\theta^\alpha
+\frac14\omega_\mu^{ab}\left(\gamma_{ab}\right)^\alpha{}_\beta\theta^\beta=\partial_\mu\theta^\alpha
~,}
where in the last equality we took \eqref{sdi}, \eqref{sdom} into account. It follows in particular that covariantly-constant negative-chirality spinors are just constant. 
We may therefore choose their basis $\theta_{(1)}^\alpha$, $\theta_{(2)}^\alpha$ as follows, in the chiral gamma-matrix basis of appendix \ref{app:spin}, 
\eq{\label{ccs}
\theta_{(1)}^\alpha=
  \left( {\begin{array}{c}
  1  \\
  0 
  \end{array} } \right)~;~~~
  \theta_{(2)}^\alpha=
  \left( {\begin{array}{c}
  0  \\
  1 
  \end{array} } \right)
~.}
The Atiyah-Patodi-Singer theorem for ALE spaces predicts an equal number of positive- and negative-chirality spinor zeromodes for the Dirac operator
$\slashed{\nabla}$  \cite{Hawking:1979zs}. On the other hand we have,
\eq{\spl{
\big(\slashed{\nabla}^{2}\theta\big)^\alpha
&=\big(\nabla^2\theta+\gamma^{\mu\nu}\nabla_\mu\nabla_\nu\theta\big)^\alpha\\
&=\big(\nabla^2\theta
+\frac18 R_{\mu\nu\rho\sigma}  \gamma^{\mu\nu} 
 \gamma^{\rho\sigma} \theta\big)^\alpha\\
&=\big(\nabla^2\theta\big)^\alpha~,}}
where in the last equality  we took \eqref{rf} into account. 
It follows that negative-chirality zeromodes are (covariantly) constant, hence non-normalizable since the ALE space is  noncompact. It thus follows from the index theorem that there are no (normalizable) spinor zeromodes of the Dirac operator.

For a spin-1 field\footnote{By a ``spin-1 field'' we understand a field transforming in the three-dimensional irreducible representation of the $su(2)$ algebra. It can be thought of as a field with 
two symmetric spinor indices of the same chirality, $\phi_{\alpha\beta}=\phi_{\beta\alpha}$ (positive chirality) or $\phi^{\alpha\beta}=\phi^{\beta\alpha}$ (negative chirality).} the index theorem predicts that the number of positive-chirality zeromodes of the Dirac operator minus the number of  negative-chirality zeromodes is equal to  the Hirzebruch signature of the ALE space. We now have,
\eq{\label{fzm}
\big(\slashed{\nabla}^{2}\phi\big)_{\alpha\beta}=
\big(\nabla^2\phi\big)_{\alpha\beta}+\frac18 R_{\mu\nu\rho\sigma} \left(\gamma^{\mu\nu}\right)_{\alpha}{}^{\alpha'}  
 \left(\gamma^{\rho\sigma}\right)_{\beta}{}^{\beta'}
\phi_{\alpha'\beta'}~;~~~
\big(\slashed{\nabla}^{2}\phi\big)^{\alpha\beta}=\big(\nabla^2\phi\big)^{\alpha\beta}
~,}
where in the second equation we took \eqref{sdi} into account. By the same argument as before, it follows that $\phi^{\alpha\beta}$ is covariantly constant, hence non-renormalizable. 
Therefore there are no spin-1 fields of negative chirality. By the index theorem it follows that there are $\tau$ spin-1 zeromodes of positive chirality (i.e.~one zeromode for the EH space).

A massless gravitino $\psi_\mu$ is also a zeromode of the Dirac operator $\slashed{\nabla}$, in the gauge $\gamma^\mu\psi_\mu=0$. 
By a similar argument as before, there are $2\tau$ spin-3/2 zeromodes of positive chirality. 
These  can be constructed as follows,
\eq{\label{psiconstr}
\psi_{(i)\mu\alpha}=\phi_{\alpha\beta}\theta_{(i)}^\gamma\left(C^{-1}\gamma_\mu\right)_{\gamma}{}^\beta{}
~;~i=1,2
~,}
where $\theta_{(i)}$ are the covariantly-constant spinors of \eqref{ccs}, and $\phi_{\alpha\beta}$ are the positive-chirality spin-1 zeromodes of \eqref{fzm}. Indeed we verify that 
the $\psi_{(i)\mu\alpha}$ are traceless,
\eq{
\left(\gamma^\mu\right)^{\alpha\beta}
\psi_{(i)\mu\alpha}=0~; ~i=1,2
~,}
as follows from \eqref{psiconstr} and  the identity $\left(C^{-1}\gamma_\mu\right)_{(\gamma}{}^\beta
\left(C^{-1}\gamma^\mu\right)_{\delta)}{}^\alpha=0$. Moreover they obey the zeromode equation,
\eq{
\big(\slashed{\nabla}^{2}\psi_\mu\big)_{\alpha}
=\big(\nabla^2\psi_\mu
+\frac12  \gamma^{\rho\sigma} R_{\mu\rho\sigma}{}^\nu   
 \psi_\nu\big)_\alpha
=0
~,}
where we used \eqref{psiconstr} and the Hodge duality relations \eqref{hm}.

\section{Gravitino condensates in 4d $\mathcal{N}=1$ supergravity}\label{app:grcond}

Within the context of 4d $\mathcal{N}=1$ supergravity, the condensate $\langle \psi^{\mu}\psi'_{\mu}\rangle$ was shown in \cite{Hawking:1979zs} to be proportional to the zeromode bilinear. 
From \eqref{psiconstr} we get,
\eq{\label{62}
\tilde{\psi}_{(1)\mu}\psi^\mu_{(2)}=f
~;~~~
f:=2C^{\alpha\beta}\phi_{\beta\gamma}C^{\gamma\delta}\phi_{\delta\alpha}
~,}
where $f$ is a positive function on the ALE space, and we have normalized  
$\tilde{\theta}_{(1)}\theta_{(2)}=1$. In deriving the above we have noted that $\phi_{\alpha\gamma}C^{\gamma\delta}\phi_{\delta\beta}$ is antisymmetric in its free indices, 
therefore it is necessarily proportional to 
$C_{\alpha\beta}$, since there is a unique scalar in the decomposition of the antisymmetric product of two spinors of positive chirality. For the EH space, cf.~\eqref{m}, $f$ can be given explicitly as in \cite{Konishi:1988mb},
\eq{\label{63}
f=16\Big(
\frac{a}{r}\Big)^8~.}
The zeromode normalization can thus be inferred from,
\eq{\label{64}
\int \d^4x\sqrt{g}~\! \tilde{\psi}_{(1)\mu}\psi^\mu_{(2)}=
\frac12\text{Vol}(S^3)\int_a^\infty\d r~\!r^3
f=4\pi^2 a^4~,}
where the ``spherical'' coordinates in \eqref{m} are related to the cartesian coordinates $x^\mu$ in the usual way, except that  antipodal points on $S^3$ are identified, see below \eqref{52}. 

To calculate the gravitino bilinear we follow \cite{Konishi:1988mb} who adopt the prescription of \cite{Gibbons:1978ac} for the functional integration over metrics. 
As shown explicitly in \cite{Konishi:1988mb} in the case of 4d $\mathcal{N}=1$ supergravity, 
expanding  the action around the EH instanton saddle point and performing the Gaussian integrations, 
the one-loop determinants from all massive modes cancel out thanks to supersymmetry. 
One is then left with the integration over zeromodes. The latter reduces to an integration over the instanton size, 
\eq{\spl{\label{65}
\langle \tilde{\psi}_{\mu}\psi^\mu \rangle &=
\text{const.} ~\! M_{\text{P}}~\!e^{-S_0}\int\d a~\! a^5~\!\tilde{\psi}_{(1)\mu}\psi^\mu_{(2)} \\
&=\text{const.} ~\! M_{\text{P}}~\! e^{-S_0}\int\d a~\! a^5 \Big(
\frac{a}{r}\Big)^8 a^{-4}
~,}}
where we have used \eqref{62}, \eqref{63} and normalized $\psi_\mu\rightarrow \psi_\mu /(2\pi a^2)$, cf.~\eqref{64}; the remaining power of $a$ comes from the Jacobian 
of the transformation from the integration over metric zeromodes to the integration over the instanton moduli. 

The integration in \eqref{65} would seem to depend on the spacetime position, since $a$ is bounded above by the radial distance $r$. 
In order to overcome this problem,   \cite{Konishi:1988mb}  performs a coordinate transformation,
\eq{\label{668}
\tilde{x}^\mu=\frac{u}{r}x^\mu~;~~~u:=r\sqrt{1-\big(\tfrac{a}{r}\big)^4}
~,}
which has the effect of changing the  radial coordinate from $r\geq a$ to $u\geq 0$. 
We can then rewrite \eqref{65} as follows,
\eq{\label{67}
\langle \tilde{\psi}_{\mu}\psi^\mu \rangle =\text{const.} ~\! M_{\text{P}}~\! e^{-S_0} \int_0^\infty\d a~\! a^9 \big(
u^2  +\sqrt{4a^4+u^4}\big)^{-4} 
~.}
This integral diverges for $a\rightarrow\infty$ at fixed $u$. In contrast, the same calculation for 
the gravitino fieldstrength bilinear $\langle (\nabla_{[\mu}{\psi}_{\nu]})^2\rangle $ 
yields a finite result \cite{Konishi:1988mb}. This is due to the fact that the two derivatives bring about an extra 
$(u/r^2)^2$ factor compared to the integrand in \eqref{67}, which contributes an extra $a^{-4}$ factor in the $a\rightarrow\infty$ limit. 
However even this finite result seems to rely on the coordinate system \eqref{668}. This does not seem  satisfactory: for 
diffeomorphism invariance to be respected, the result should be independent of the coordinate system used for its calculation.

One may argue that  the divergence/ambiguity encountered is not    surprising 
since the 4d theory is  nonrenormalizable and should anyway be thought of as an effective low-energy limit of string theory. 
On general grounds,  at one loop in the gravitational coupling, one expects a 
gravitational instanton contribution to the 
condensate of the form,
\eq{\label{68}
\langle \tilde{\psi}_{\mu}\psi^\mu \rangle \propto  M_{\text{P}}~\!e^{-S_0}\propto   l_s^{-1}e^{-c~\!(l_Y/l_s)^2}
~,}
up to proportionality constants of order one,  where in the second proportionality  we have taken \eqref{8bc}, \eqref{s01} into account. 
Similarly, the quartic gravitino condensate receives contributions from the ALE with $\tau=2$ and scales as the square of the 
bilinear condensate above.

%
%

\end{document}